\documentclass[aps,preprint,amsmath,tightenlines,showpacs,showkeys]{revtex4}

\usepackage{bm}
\usepackage{graphicx}
\usepackage{epsfig}

\begin{document}

\def\US{\Upsilon}
\def\zb{Z^{\pm}_b}
\def\zbp{Z^{'\pm}_b}

\title{Decay coupling constant sum rules for tetraquarks
$\bm{T[(\bar{Q}q)(Q\bar{q})]}$ with broken SU(3) symmetry}

\author{V.~Gupta}

\email[]{virendra@mda.cinvestav.mx}

\affiliation{
Departamento de F\'{\i}sica Aplicada.\\
Centro de Investigaci\'on y de Estudios Avanzados del IPN.\\
Unidad M\'erida.\\
A.P. 73, Cordemex.\\
M\'erida, Yucat\'an, 97310. MEXICO.
}

\author{G.~S\'anchez-Col\'on}

\email[]{gsanchez@mda.cinvestav.mx}

\affiliation{
Departamento de F\'{\i}sica Aplicada.\\
Centro de Investigaci\'on y de Estudios Avanzados del IPN.\\
Unidad M\'erida.\\
A.P. 73, Cordemex.\\
M\'erida, Yucat\'an, 97310. MEXICO.
}

\author{S.~Rajpoot}

\email[]{rajpoot@csulb.edu}

\affiliation{
Department of Physics \& Astronomy.\\
California State University, Long Beach.\\
Long Beach, CA 90840. USA.
}

\date{\today}

\begin{abstract}

For tetraquarks of the form $T[(\bar{Q}q)(Q\bar{q})]$ we give sum rules for
their decay coupling constants, taking into account the SU(3) symmetry breaking
interactions to first order.

\end{abstract}



\maketitle

\section{\label{introduction}Introduction}

Two hidden-bottom charged meson resonances, $Z_b(10610)$ and $Z_b(10650)$ with
$J^P=1^+$, were observed recently by the Belle Collaboration~\cite{belle12}.
Since these decay into $\pi^{\pm}\Upsilon(nS)$ ($n=1,2,3$) and
$\pi^{\pm}h_b(mP)$ ($m=1,2$) they have been
interpretated~\cite{Guo,Richard,gupta05} as tetraquarks of the form
$T(Q\bar{Q}\,q\bar{q})$. A natural decay mode for such a tetraquark would be

\begin{equation}
T(Q\bar{Q}\,q\bar{q})\to H(Q\bar{Q}) +  M(q\bar{q}),
\label{tetraquark}
\end{equation}

\noindent
where $H(Q\bar{Q})$ is a heavy meson with $Q=b\ {\rm or}\ c$ and
$M(q\bar{q})$ could be the pseudoscalar or vector meson SU(3) flavour octet.
For such tetraquarks we recently~\cite{gupta12} gave sum rules for tetraquark
decay coupling constants taking into account the SU(3) breaking interactions to
first order.

In this paper we consider tetraquarks of the form $T[(\bar{Q}q)\,(Q\bar{q})]$
where the heavy quark $Q=b\ {\rm or}\ c$ and the light quarks $q=u,d,\ {\rm
or}\ s$. For a given $Q$ there will be nine such tetraquarks. The natural
decay mode will be into two mesons $(\bar{Q}q)$ and $(Q\bar{q})$ subject to
angular momentum and parity selection rules. For example, a $J^P=0^+$
tetraquark would decay into two pseudoscalar mesons.

We consider the decays

\begin{equation}
T[(\bar{Q}q)\,(Q\bar{q})]\to M(\bar{Q}q) +  M(Q\bar{q}),
\label{tetraquarkdos}
\end{equation}

\noindent
in Sec.~\ref{sumrules} below. We give sum rules for the nine decay
couplings constants taking into account the SU(3) breaking interactions to
first order.

An important difference between the tetraquarks $T(Q\bar{Q}\,q\bar{q})$
considered earlier and the tetraquarks $T[(\bar{Q}q)\,(Q\bar{q})]$ considered
here is that the latter can have OZI~\cite{ozi1,ozi2,ozi3} suppressed decays
through annihilation of $\bar{Q}^{\alpha}Q_{\beta}$ through gluons into lighter
quarks $\bar{q}^{\alpha}q_{\beta}$ (greek letters represent the color indices).
These modes are briefly discussed in the concluding remarks.

\section{\label{notation}Notation}

We denote the tetraquark made of a heavy meson and its antiparticle as

\begin{equation}
T^i_k = T[(\bar{Q}q_k)\,(Q\bar{q}^i)],
\label{tetraquarktres}
\end{equation}

\noindent
with $i,k=1,2,3$ (latin letters representing flavour indices). The
heavy quark $Q=b\ {\rm or}\ c$ is SU(3) flavour singlet. The light quarks $q_k$
($k=1,2,3$) transform as SU(3) $\bm{3}$ while the antiquarks $\bar{q}^i$
($i=1,2,3$) transform as $\bar{\bm{3}}$. In color space,
$M_k\equiv(\bar{Q}q_k)$ and $\bar{M}^i\equiv(Q\bar{q}^i)$ form color singlets
and transform as $\bm{3}$ and $\bar{\bm{3}}$ of SU(3) flavour, respectively.

The natural decay mode for the tetraquark would be

\begin{equation}
T^i_k \to M_k+\bar{M}^i.
\label{decaymode}
\end{equation}

\noindent
The tetraquark states can be represented as a $3\times 3$ matrix $T$
with $T^i_k$ as matrix elements, so that,

\begin{eqnarray}
T&=&\left(\begin{array}{ccc}
T^1_1 & T^2_1 & T^3_1 \\
T^1_2 & T^2_2 & T^3_2 \\
T^1_3 & T^2_3 & T^3_3
\end{array}\right).
\label{tmatriz}
\end{eqnarray}

\noindent In SU(3) flavour space this transforms as $\bm{1}\oplus\bm{8}$. The
mesons $M_k$ and $\bar{M}^i$ transform as SU(3) $\bm{3}$ and $\bar{\bm{3}}$,
respectively. The $3\times 3$ matrix $\mathcal{M}$ representing the final states
is

\begin{eqnarray}
\mathcal{M}&=&\left(\begin{array}{ccc}
\mathcal{M}^1_1 & \mathcal{M}^2_1 & \mathcal{M}^3_1 \\
\mathcal{M}^1_2 & \mathcal{M}^2_2 & \mathcal{M}^3_2 \\
\mathcal{M}^1_3 & \mathcal{M}^2_3 & \mathcal{M}^3_3
\end{array}\right),
\label{mmatriz}
\end{eqnarray}

\noindent
where $\mathcal{M}^i_k$ represents $M_k+\bar{M}^i$ in the final state. In
flavour space it also transforms as $\bm{1}\oplus\bm{8}$, this is clear since
$(\bar{Q}q)\otimes(Q\bar{q})$ transform as $\bm{3}\otimes
\bar{\bm{3}}=\bm{1}\oplus\bm{8}$.

\section{\label{sumrules}Sum rules}

In unbroken SU(3) there will be a single decay coupling constant $G_0$; with
$\lambda_8$ breaking the tetraquark octet can form two octets, symmetric
(coupling constant $G_D$) and antisymmetric (coupling constant $G_F$). In
matrix form we can write the decay coupling constant as

\begin{eqnarray}
\lefteqn{G\left[T^i_k(\bm{8})\to M_k+\bar{M}^i\right]=}\nonumber\\
 & & G_0\,{\rm Tr}[\widetilde{T}\mathcal{M}] +
G_D\,{\rm Tr}[\widetilde{T}(\lambda_8\mathcal{M}+\mathcal{M}\lambda_8)]+
G_F\,{\rm Tr}[\widetilde{T}(\lambda_8\mathcal{M}-\mathcal{M}\lambda_8)].
\label{sumrulet}
\end{eqnarray}

\noindent
Explicitly, for the nine tetraquark decays

\begin{equation}
G(T^1_1 \to M_1+\bar{M}^1)=G_0+2G_D,
\label{uno}
\end{equation}

\begin{equation}
G(T^2_2 \to M_2+\bar{M}^2)=G_0+2G_D,
\label{dos}
\end{equation}

\begin{equation}
G(T^3_3 \to M_3+\bar{M}^3)=G_0-4G_D,
\label{tres}
\end{equation}

\begin{equation}
G(T^2_1 \to M_1+\bar{M}^2)=G_0+2G_D,
\label{cuatro}
\end{equation}

\begin{equation}
G(T^1_2 \to M_2+\bar{M}^1)=G_0+2G_D,
\label{cinco}
\end{equation}

\begin{equation}
G(T^3_1 \to M_1+\bar{M}^3)=G_0-G_D+3G_F,
\label{seis}
\end{equation}

\begin{equation}
G(T^3_2 \to M_2+\bar{M}^3)=G_0-G_D+3G_F,
\label{siete}
\end{equation}

\begin{equation}
G(T^1_3 \to M_3+\bar{M}^1)=G_0-G_D-3G_F,
\label{ocho}
\end{equation}

\begin{equation}
G(T^2_3 \to M_3+\bar{M}^2)=G_0-G_D-3G_F.
\label{nueve}
\end{equation}

\noindent
Nine decays and three coupling constants. So, six sum rules or relations,

\begin{equation}
G(T^1_1 \to M_1+\bar{M}^1)=G(T^2_2 \to M_2+\bar{M}^2)=
G(T^2_1 \to M_1+\bar{M}^2)=G(T^1_2 \to M_2+\bar{M}^1);
\label{ruleunoatres}
\end{equation}

\begin{equation}
G(T^3_1 \to M_1+\bar{M}^3)=G(T^3_2 \to M_2+\bar{M}^3),\quad
G(T^1_3 \to M_3+\bar{M}^1)=G(T^2_3 \to M_3+\bar{M}^2);
\label{rulecuatrocinco}
\end{equation}

\begin{equation}
G(T^2_2 \to M_2+\bar{M}^2)+G(T^3_3 \to M_3+\bar{M}^3)=
G(T^3_1 \to M_1+\bar{M}^3)+G(T^1_3 \to M_3+\bar{M}^1).
\label{ruleseis}
\end{equation}

If and when such tetraquarks are observed one can extract the coupling constants
from the observed decay rates. For a $J^P=0^+$ tetraquark decaying into two
pseudoscalar mesons the decay width is

\begin{equation}
\Gamma=\frac{1}{8\pi}\frac{k}{M^2}|A|^2,
\end{equation}

\noindent
where $M$ is the tetraquark mass, $k$ is the momentum of a decay
particle and $A$ is the transition amplitude. For $s$-wave decay ($0^+\to
0^- + 0^-$) $A=GM$ with $G$ the coupling constant.

\section{\label{conclusions}Concluding remarks}

\begin{figure}
\centerline{\psfig{file=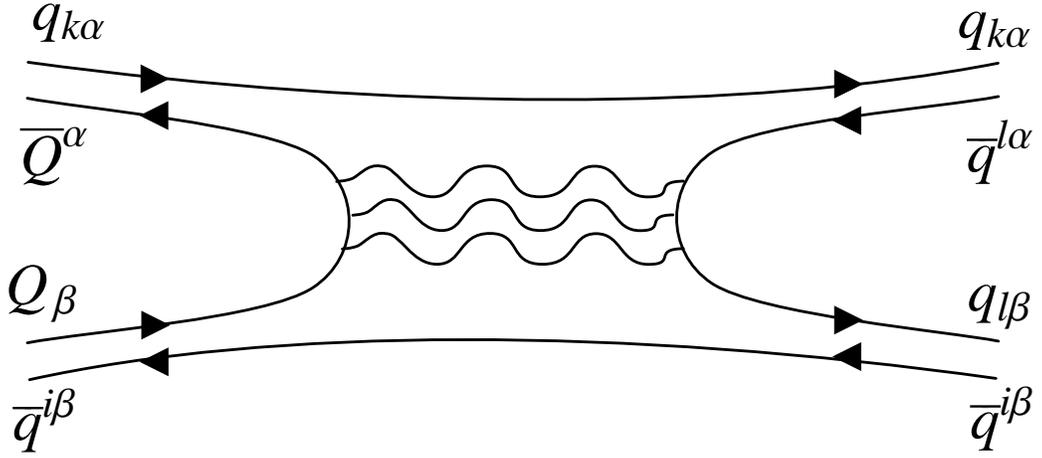,width=6.4in}}
\caption{
\label{figure}
Tetraquarks $T[(\bar{Q}q)\,(Q\bar{q})]$ observed through their OZI suppressed
decay into light meson plus anti meson pairs, $q_k\bar{q}^l + q_l\bar{q}^i$.
Greek letters represent the color indices and latin letters represent flavor
indices.
}
\end{figure}

To date the tetraquarks $T[(\bar{Q}q)\,(Q\bar{q})]$ have not been observed
experimentally. However, their decays products, the heavy mesons (for example
$D^+_S = c\bar{s}$, $B^+ = u\bar{b}$, etc.) have been observed~\cite{pdg12}. It
is conceivable that this type of tetraquark may be seen as a resonance in the
heavy quark particle-antiparticle mass spectra, for example, $D^+_S + D^-_S$,
$D^0 + \bar{D}^0$, etc. Another possibility is that they may be observed
through their decay into light meson plus anti meson pairs, for example,
$q_k\bar{q}^l + q_l\bar{q}^i$ (see Fig.~\ref{figure}). These are possible
through annihilation of $\bar{Q}Q$ through gluons into $\bar{q}q$. However,
such decays are likely to be OZI~\cite{ozi1,ozi2,ozi3} suppressed.

\begin{acknowledgments}

V.~Gupta and G.~S\'anchez-Col\'on would like to thank CONACyT (M\'exico) for
partial support. The work of S.~Rajpoot was supported by DOE Grant \#:
{DE-FG02-10ER41693}.

\end{acknowledgments}

\clearpage

\begin{thebibliography}{}

\bibitem{belle12}

Belle~Collab. (A.~Bondar {\it et al.}), {\it Phys.\ Rev.\
Lett.} {\bf 108}, 122001 (2012).

\bibitem{Guo}

T.~Guo, L.~Cao, M.~Z.~Zhou and H.~Chen, arXiv: 1106.2284 [hep-ph].

\bibitem{Richard}

F.~S.~Navarra, M.~Nielsen and J.-~M.~Richard, arXiv: 1108.1230 [hep-ph].

\bibitem{gupta05}

For a discussion of all tetraquarks containing only $c$ and $s$ quarks see:
V.~Gupta, {\it Int.\ J.\ Mod.\ Phys.\ A.} {\bf 20}, 5891 (2005).

\bibitem{gupta12}

V.~Gupta, G.~S\'anchez-Col\'on, and S.~Rajpoot, {\it Mod.\ Phys.\ Lett.\ A}
{\bf 27}, 1250165 (2012).

\bibitem{ozi1}

S.~Okubo, {\it Phys.\ Lett.} {\bf 5}, 165 (1963).

\bibitem{ozi2}

G.~Zweig, CERN\ Report.\ TH 401 and 412 (1964).

\bibitem{ozi3}

J.~Izuka, {\it Prog.\ Theor.\ Phys.\ Suppl.} {\bf 37} and {\bf 38}, 21 (1966).

\bibitem{pdg12}

Particle Data Group (J.~Beringer {\it et al.}), {\it Phys.\ Rev.\ D} {\bf 86},
010001 (2012).

\end{thebibliography}
\end{document}